# From “Hallmarks of Cancer” to “Passive Survival System”: A Paradigmatic Reconstruction of Oncological Theory through the Existential Crisis–Driven Survival (ECDS) Framework

Yuxuan Zhang[1], Lijun Jia[1]*

[1]School of Integrative Medicine, Nanjing University of Chinese Medicine, Nanjing 210023, China

*Corresponding author: Lijun Jia, E-mail: jlj@njucm.edu.cn; jialijun2002@aliyun.com

**Author contributions**: Lijun Jia: Supervised the project, conceptualized the Existential Crisis-Driven Survival (ECDS) theoretical framework, integrated classical oncology theories, and critically revised the manuscript. Yuxuan Zhang: Led manuscript drafting, refined the theoretical logic, and developed academic schematics to illustrate the conceptual architecture.

**Competing interests**: The authors declare no competing interests.

**Abstract:**

Malignant tumors exhibit complex pathogenesis, yet classical oncological theories remain fragmented, failing to provide a unifying framework to address this complexity. This gap limits the utility and translational potential of the prevailing "confront-and-eradicate" therapeutic paradigm, constraining transformative therapeutic breakthroughs and driving the emergence of acquired and recurrent drug resistance. Here, we propose the Tumor Existential Crisis-Driven Survival (ECDS) theory, anchored in the core proposition that impairment of Existential Stability drives the compensatory hyperactivation of Survival Capacity. This framework defines three foundational constructs (Existential Stability, Survival Capacity, and Existence Threshold) and three guiding principles, unifying and integrating canonical core theories of tumorigenesis. It delineates the dynamic coupling between declining Existential Stability and escalating Survival Capacity during tumor evolution, reinterprets the hierarchical activation of the well-established 14 cancer hallmarks, elucidates the redundancy of survival signaling pathways that underpins intratumoral and intertumoral heterogeneity, and unravels the "hierarchical leap" in therapeutic resistance. By reframing tumors as "Existential Stability erosion-driven passive survival systems" rather than "intrinsically aggressive cellular aggregates", ECDS challenges prevailing dogma, uncovers tumors' intrinsic vulnerability, and establishes a robust meta-theoretical foundation for both basic cancer research and translational clinical management.

## 1. Introduction

Malignant tumors constitute one of the most pressing global health challenges, distinguished by their etiological complexity and persistent clinical intractability.[1,2] Over the past five decades, landmark conceptual advances—including the urvival stabilitysomatic mutation theory and the “hallmarks of cancer” framework—have profoundly shaped our understanding of carcinogenesis and enabled the development of life-prolonging targeted and immunotherapies.[3-6] Nevertheless, classical paradigms suffer from pronounced conceptual compartmentalization: the somatic mutation theory emphasizes initiating genetic lesions,[7,8] clonal evolution theory focuses on Darwinian selection within heterogeneous cell populations,[9,10] and microenvironmental theory centers on bidirectional crosstalk between tumor cells and stromal components.[11] Collectively, these frameworks fail to provide a mechanistic bridge across biological scales—from molecular perturbation to tissue-level behavior—and across temporal stages—from initiation to metastasis and relapse. Clinically, this fragmentation manifests in persistent challenges: profound inter- and intra-tumoral heterogeneity, rapid selection of therapy-resistant subclones, and systemic toxicity limiting therapeutic index.[12]

These enduring limitations underscore the necessity of moving beyond descriptive phenomenology toward a predictive, systems-level theory grounded in universal evolutionary principles. Accordingly, oncology requires three interdependent paradigm shifts: (i) recognition of overarching organizing principles governing tumor behavior; (ii) acknowledgment of systemic, multi-scale regulatory logic—not merely linear causality; and (iii) integration of disparate theories into a coherent, hierarchically structured meta-framework. To address this need, we introduce the Existential Crisis–Driven Survival (ECDS) theory. It posits that malignancy emerges not from autonomous aggressiveness, but from the irreversible erosion of Existential Stability (Φ), compelling tumor cells to deploy successive, energetically costly, and increasingly maladaptive survival strategies. Underlying the apparent phenotypic aggression of tumors lies a fundamental “intrinsic fragility”: driver mutations, immune evasion, angiogenesis, and metastatic dissemination are not intrinsic drivers of progression but rather context-dependent, compensatory adaptations to deteriorating Φ—each entailing substantial fitness trade-offs.

This article systematically develops the ECDS theoretical architecture: it formalizes its core constructs and principles; integrates classical models of tumor initiation and progression; interprets multistage tumor evolution through the lens of Φ–Ψ dynamics; recontextualizes the canonical hallmarks as sequential, threshold-governed manifestations of Ψ activation; analyzes heterogeneity and therapeutic resistance as emergent properties of survival-pathway redundancy and hierarchical adaptation; and outlines translational implications for diagnostics, therapeutic design, and longitudinal clinical management. Ultimately, ECDS catalyzes a foundational shift—from "confrontational medicine," predicated on direct cytotoxic elimination, to "system-regulated medicine," guided by modulation of the Φ–Ψ equilibrium. Its unifying causal logic—genetic and epigenetic perturbations → Φ impairment → synergistic amplification by microenvironmental stressors → Ψ activation → transient survival or futile adaptation—provides a comprehensive, evolutionarily grounded explanatory scaffold for the full spectrum of malignant behavior, thereby establishing a transformative meta-theoretical foundation for 21st-century oncology.

## 2. Core Concepts and Principles of the ECDS Theory

### 2.1 Qualitative Definition of Core Concepts

Within the ECDS framework, tumor cells are formally conceptualized as autonomous biological subsystems governed by an intrinsic imperative of self-persistence. To rigorously characterize their persistence states and evolutionary trajectories, the theory introduces three foundational, interdefined concepts.

#### 2.1.1 Existential Stability (Φ): A Multidimensional Metric of Cellular Persistence Capacity

Φ quantifies a cell's intrinsic capacity to sustain its existence across three functionally integrated dimensions: survival stability, survival duration, and survival scope (Figure 1). Each dimension is anchored in empirically measurable biological parameters:

- Survival stability: Defined hierarchically by multiple major mutually reinforcing axes, including genomic integrity (e.g., DNA repair fidelity, chromosomal segregation accuracy) → metabolic homeostasis (e.g., mitochondrial respiratory efficiency, redox balance via ROS scavenging systems) → microenvironmental adaptability (e.g., dynamic responsiveness to immune surveillance, stromal adhesion plasticity, vascular niche integration).[13,14]
- Survival duration: Governed primarily by multiple major regulatory factors, with telomere maintenance mechanisms—including telomerase reactivation and the alternative lengthening of telomeres (ALT) pathway—and the functional integrity of senescence-inducing signaling cascades (e.g., p53–p21 and p16–RB pathways) acting as key contributors.[15,16]

- Survival scope: Reflects the degree of dependence on extrinsic microenvironmental cues (e.g., growth factor availability, cell–cell and cell–matrix adhesion signals); a narrower survival scope signifies heightened contextual dependency and directly arises from compromised microenvironmental adaptability.[17]

Endogenous or exogenous stressors—including driver mutations, chronic inflammation, hypoxia, and genotoxic exposures—induce coordinated, synchronous erosion across all three dimensions of Φ, thereby undermining the integrated physiological competence required for homeostatic cellular function.[18] Among these, genetic variation constitutes the primary genotypic trigger of Φ impairment: only canonical driver mutations (e.g., activating KRAS mutations or biallelic TP53 loss) directly disrupt core Φ dimensions; passenger mutations accumulate passively as epiphenomena without causal contribution to Φ decline. Non-genetic stressors—such as chronic inflammation—typically act not as initiators but as synergistic amplifiers, either accelerating the accumulation of genetic lesions or lowering the activation threshold for compensatory Survival Capacity (Ψ) responses.

Critically, Φ impairment in malignant cells is unidirectional and irreversible—a defining feature that distinguishes neoplastic transformation from transient physiological stress responses. This irreversibility stems from the collapse of core maintenance mechanisms: tumor cells lack the functional redundancy and regulatory coordination necessary to restore genomic integrity, metabolic equilibrium, or microenvironmental fidelity through endogenous repair. In contrast, normal somatic cells retain robust homeostatic capacity; mild Φ perturbations (e.g., low-level oxidative DNA damage) are effectively resolved via coordinated upregulation of homologous recombination, antioxidant defenses, and epithelial-mesenchymal signal normalization—thereby preserving Φ without engaging pathological compensation. The rate and magnitude of Φ erosion are thus predictive biomarkers of malignant risk, correlating with both the tempo of carcinogenesis and clinical outcomes in a gradient-dependent manner.

### 2.1.2 Survival Capacity (Ψ): A Context-Dependent Phenotypic Compensation Mechanism

Ψ denotes the cell's capacity to deploy evolutionarily co-opted, non-physiological survival programs in response to Φ impairment. Rather than representing an intrinsic aggressive drive, Ψ activation is a passive, adaptive phenotypic response to functional deficits induced by driver mutations: proto-oncogene activation or tumor suppressor inactivation compromises Φ, prompting compensatory engagement of Ψ-associated pathways—modulated and potentiated by microenvironmental cues (Figure 1). A fundamental distinction exists in the Ψ activation threshold between normal and transformed cells: under equivalent Φ impairment (e.g., sublethal oxidative DNA damage), normal cells maintain viability solely through physiological repair and signal restoration, without invoking Ψ; tumor cells—characterized by constitutively low baseline Φ and defective core maintenance machinery—exhibit markedly lowered Ψ thresholds and rapidly engage high-amplitude compensatory responses (e.g., Bcl-2 family overexpression, PI3K–AKT hyperactivation), with intensity orders of magnitude exceeding that observed in non-transformed counterparts.[19]

Ψ-mediated survival exhibits three defining properties that underpin tumor evolution and therapeutic resistance:[3-5]

- **Multidimensionality:** Activation spans functionally distinct yet complementary axes—including dysregulated proliferation, apoptosis evasion, metabolic reprogramming, immune escape and invasion–metastasis competence—each specifically counteracting a corresponding deficit in a particular Φ dimension;
- **Multi-pathway redundancy:** Single survival objectives (e.g., resistance to cell death) are achieved through parallel, mechanistically diverse strategies, conferring robustness against therapeutic perturbation or microenvironmental fluctuation;
- **Progressive upgradability:** As Φ deteriorates, Ψ responses escalate hierarchically—from baseline proliferative autonomy, to immune evasion, and ultimately to invasive and metastatic phenotypes—with each transition entailing increased molecular complexity, energetic cost, and systemic dependency.

Importantly, each advancement in Ψ activity exacerbates further Φ erosion: increasingly complex survival behaviors heighten cellular reliance on extrinsic nutrients, growth factors, and stromal support, thereby diminishing intrinsic survival stability and establishing a self-reinforcing feedback loop of Φ decline and Ψ hyperactivation.

#### 2.1.3 Existence Threshold (Θ): The Minimal Persistence Requirement for Cellular Viability

Θ represents the quantitative lower bound of integrated persistence capacity required to sustain essential cellular viability functions. It serves as the formal, mathematically tractable constraint linking Φ and Ψ through the governing principle: $\Phi + \Psi \geq \Theta$. Consistent with this governing principle, in healthy tissues,

Θ is primarily fulfilled by endogenous Φ, with physiological compensatory mechanisms providing supplementary support when minor perturbations to Φ arise—ensuring the maintenance of cellular viability without triggering pathological adaptive responses; tumor cells—facing persistent, non-compensable Φ deficits—must continuously elevate Ψ to bridge the gap and satisfy Θ. Notably, Θ is neither universal nor static: it exhibits tissue-specific calibration (e.g., higher in stem/progenitor compartments due to regenerative demands; lower in post-mitotic neurons optimized for functional longevity) and evolutionary stratification across cell lineages.[20]

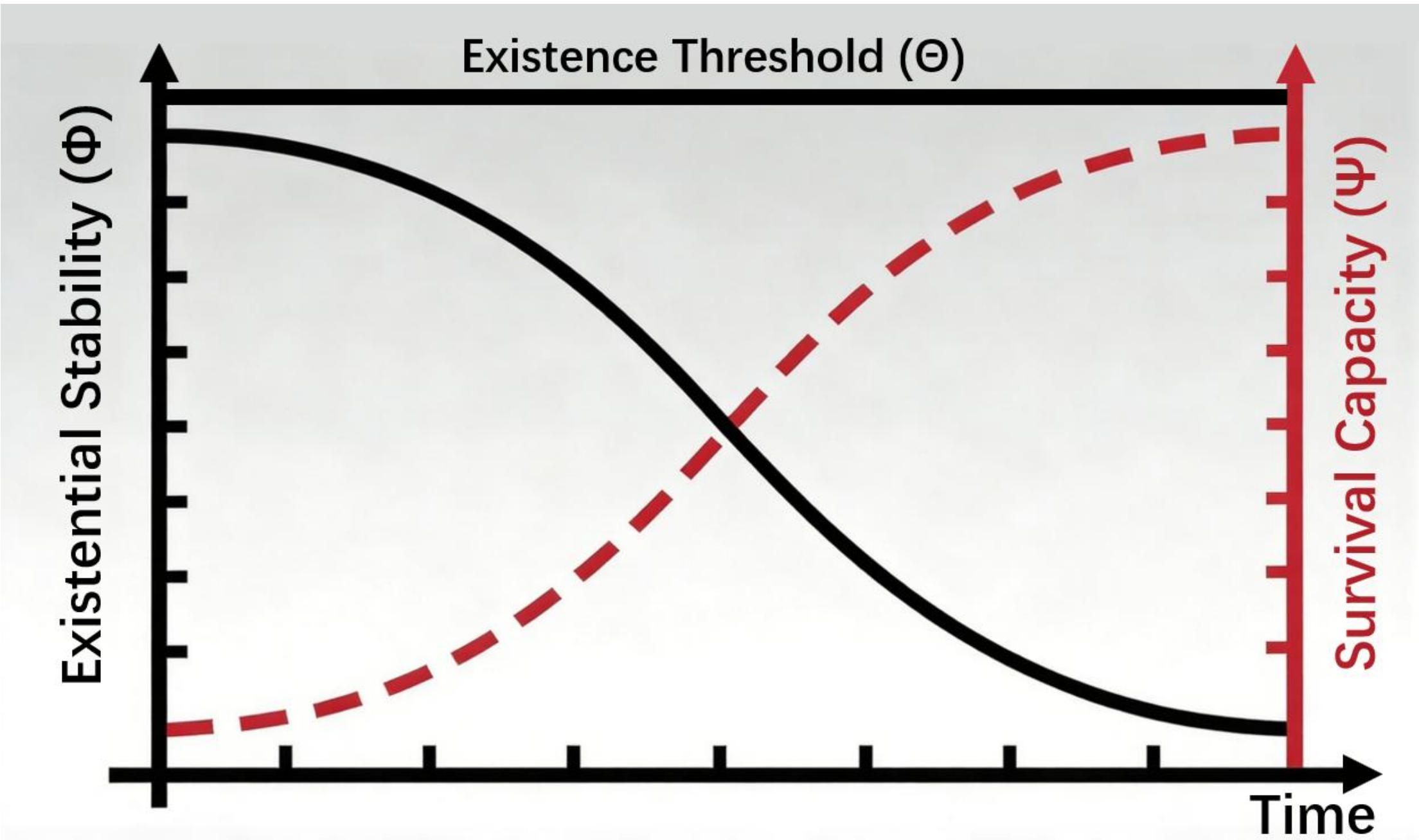

**Figure 1. Dynamic coupling of Existential Stability (Φ), Survival Capacity (Ψ), and Existence Threshold (Θ) over time**

Φ undergoes unidirectional, irreversible decline driven primarily by driver mutations (core etiologic trigger) and secondarily amplified by non-genetic stressors (e.g., inflammation, hypoxia). Ψ is passively engaged as a phenotypic compensation mechanism for Φ deficits, with its magnitude and kinetics modulated by microenvironmental context. Θ defines the minimal persistence threshold: cellular viability is sustained only when the summed contributions of Φ and Ψ meet or exceed Θ. While normal cells rely predominantly on intact Φ with minimal Ψ involvement, tumor cells depend on sustained Ψ upregulation to offset progressive Φ erosion. Critically, stable Φ–Ψ equilibrium—manifesting clinically as tumor dormancy—is achievable under conditions of mild Φ impairment or effective therapeutic modulation, demonstrating that escalation into a vicious cycle is not inevitable but context-contingent.

### 2.2 Core Principles of the ECDS Theory

Based on the interdependent definitions of Φ, Ψ, and Θ, the ECDS Theory articulates three interlocking principles that govern malignant progression (Figure 2) and collectively constitute a closed-loop theoretical architecture—capable of explaining tumor evolution across its full continuum, from initiation and clonal expansion to metastatic dissemination and systemic collapse.

#### 2.2.1 Principle of Unidirectional Existential Stability Impairment

The Existential Stability (Φ) of tumor cells undergoes progressive, irreversible, and unidirectional erosion[21]—a process driven by the convergence of two complementary forces[22,23]: (i) intrinsic evolutionary constraints inherent to multicellular organization (where somatic cell lineages sacrifice autonomous persistence capacity in exchange for tissue-specific functional specialization), and (ii) acquired microenvironmental stressors (with driver mutations serving as the primary genotypic trigger, and chronic inflammation, ionizing radiation, replicative aging, and metabolic stress acting as synergistic amplifiers). Critically, this impairment is non-recoverable in neoplastic cells and in non-malignant cells with profound DNA repair deficiencies; physiological compensation remains feasible only in normal cells experiencing mild, transient Φ perturbations. The tempo and extent of Φ erosion directly govern oncogenic trajectory: accelerated or deeper decline intensifies selective pressure for survival adaptation, thereby promoting tumor initiation, phenotypic diversification, metastatic competence, intratumoral heterogeneity, and therapeutic resistance—establishing Φ erosion rate as a quantitative determinant of clinical aggressiveness.

#### 2.2.2 Principle of Passive Adaptive Upregulation of Survival Capacity

The compensatory upregulation of Survival Capacity (Ψ) constitutes the principal phenotypic response to Φ decline. As Φ deteriorates, cells passively engage Ψ pathways—not as an autonomous program but as a homeostatic imperative—to preserve the viability constraint $\Phi + \Psi \geq \Theta$. This dynamic underlies the emergence, functional diversification, and hierarchical escalation of malignant traits. Ψ activation follows a rigorously ordered, threshold-governed progression: a basal layer (e.g., Ras–MAPK–driven hyperproliferation) → an enabling layer (e.g., VEGF–mediated angiogenesis supporting nutrient supply) → an advanced layer (e.g., Snail/Twist–induced epithelial–mesenchymal transition (EMT) and metastatic dissemination). Each adaptive tier represents a context-dependent, energetically costly solution to escalating Φ deficits—and each, in turn, imposes additional functional burdens that further destabilize Φ, thereby reinforcing a self-amplifying cycle of deterioration.[24,25]

### 2.2.3 Principle of Survival Futility

Ψ upregulation sustains cellular viability only transiently and conditionally—it cannot restore or reverse underlying Φ impairment. While progressive Φ erosion necessitates successive rounds of Ψ activation, increasingly complex survival adaptations (e.g., EMT, anoikis resistance, stromal reprogramming, immunosuppressive niche formation) incur substantial biological liabilities: heightened metabolic demand, genomic instability, loss of functional differentiation, and systemic resource depletion.[26] Consequently, each compensatory strategy exacerbates Φ erosion, entrenching the core feedback loop: Φ impairment → Ψ activation → further Φ exhaustion. Ultimately, excessive reliance on maladaptive survival mechanisms precipitates intrinsic tumor cell dysfunction, microenvironmental dysregulation, and host systemic toxicity—culminating in failure to satisfy $\Phi + \Psi \geq \Theta$ and resulting in dual extinction of both tumor and host.[27,28] Notably, this cascade is not deterministic but context-modulated: under conditions of mild Φ impairment (e.g., early-stage preinvasive lesions) or effective external regulation (e.g., targeted inhibition of key Ψ effectors, immune checkpoint blockade, or microenvironment normalization), Ψ activity may be suppressed or maintained at subthreshold levels—enabling a stable, low-energy "pathological equilibrium" manifesting clinically as tumor dormancy or long-term disease stabilization in advanced cancer patients.[29,30]

Collectively, the dynamic tripartite coupling of Φ, Ψ, and Θ forms the theoretical skeleton of the ECDS framework—providing a unified, evolutionarily coherent logic for integrating historically disparate tumor theories and systematically analyzing malignant behavior across the entire ontogenetic lifespan of cancer.

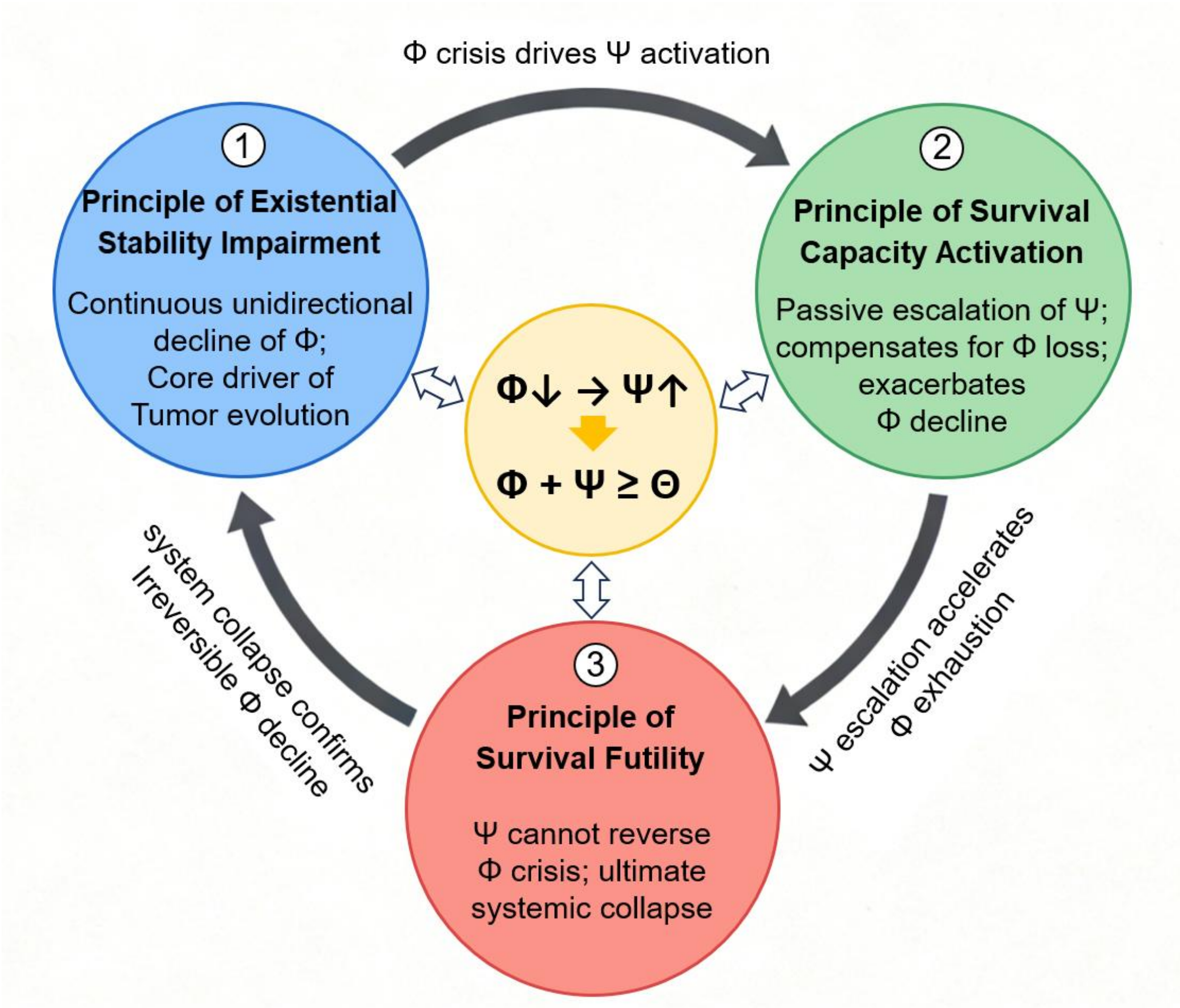

**Figure 2. Schematic illustration of the core logic and principles of the ECDS Theory**

This diagram synthesizes the central pathogenic mechanism—driver mutation–initiated Φ crisis, amplified synergistically by non-genetic stressors, leading to passive Ψ activation—and delineates the three foundational principles: (1) Continuous, unidirectional Φ erosion as the ultimate etiologic driver of tumorigenesis; (2) Passive, hierarchical Ψ escalation as a compensatory response that paradoxically accelerates Φ decline; and (3) Inherent futility of Ψ-mediated survival, culminating in systemic collapse when Φ + Ψ falls below Θ. The self-reinforcing cycle Φ impairment → Ψ escalation → Φ exhaustion constitutes the core engine of tumor progression and therapeutic failure.

## 3. Integrative Analysis of Classical Tumor Origin Theories via the ECDS Framework

Tumor origin is a multistage, nonlinear process shaped by intrinsic genetic programs and extrinsic environmental pressures.[8] Classical theories address key facets of carcinogenesis but remain siloed, failing to resolve clinical paradoxes (e.g., driver mutation carriage without carcinogenesis). The ECDS

Theory reconciles them via a unifying causal chain: driver mutation-induced Φ impairment → non-genetic stressor amplification → functional ground-state disruption → passive Ψ activation → tumor initiation.

### 3.1 Somatic Mutation Theory: The Genotypic Trigger of Φ Impairment

Driver mutations are precise disruptors of Φ integrity (e.g., TP53 loss or APC inactivation).[31,32] Oncogenic transformation requires sequential mutational accumulation to cross the Φ impairment threshold; mutational burden alone is insufficient—residual Φ above the homeostatic threshold avoids pathological Ψ activation. This distinguishes ECDS from classical somatic mutation theory, which treats mutations as isolated triggers rather than mediators of Φ collapse.

### 3.2 "Bad Luck" Theory: The Stochastic Accelerator of Φ Erosion

Stochastic errors (DNA replication errors, epigenetic drift) during stem cell division accelerate Φ erosion—stem cells operate near Φ's lower boundary, and cumulative probabilistic lesions trigger Ψ activation only when exceeding physiological compensation. This reframes classical "bad luck" theory by embedding stochasticity within the Φ–Ψ–Θ regulatory architecture, explaining why only Φ-impairing errors culminate in malignancy.[8]

### 3.3 Tissue Field Theory: The Microenvironmental Modulator of Φ–Ψ Coupling

Classical tissue field theory (epithelial–stromal crosstalk disruptions) is reconceptualized as a Φ–Ψ coupling modulator via two mechanisms:[11,33] (1) accelerating Φ impairment via diverse factors, e.g., pro-inflammatory mediators (e.g., TNF-α, ROS) that induce oxidative DNA damage, chromosomal instability, and impaired adhesion; (2) lowering Ψ activation thresholds (paracrine factors, immune-permissive niches). This distinguishes ECDS by positioning the microenvironment as a context-dependent regulator, not an independent driver.

### 3.4 Functional Ground State Theory: The Phenotypic Carrier and Transducer of Φ Impairment

The functional ground state—defined as the homeostasis of epigenetic, transcriptional, and signaling pathways—acts as the intuitive, dynamic phenotypic carrier of Φ, while linking genetic variation to Φ impairment. Ground state stability correlates directly with Φ levels: a stable ground state is characterized by conserved epigenetic modifications, coordinated core gene expression, and steady signaling activity, which collectively maintain high Φ; in contrast, a disordered ground state leads to reduced Φ and increased cancer transformation risk.[34] The ECDS Theory extends classical functional ground state theory by anchoring ground state disruption to the progressive impairment of Φ, thereby integrating phenotypic plasticity (a key feature of the ground state) into the unified Φ–Ψ theoretical framework.[35]

### 3.5 The Unified ECDS Logical Chain of Tumor Origin

The four classical theories are not mutually exclusive but occupy distinct, complementary positions in a causally coherent ECDS logical chain of tumor origin: ECDS establishes a causally coherent chain: driver mutations (genotypic trigger) + non-genetic stressors (amplifiers) → functional ground state disorder (mediator) → gradient-dependent Φ impairment → context-dependent Ψ activation → tumor initiation (**Figure 3**). The four classical theories occupy distinct roles:

1. Somatic mutation theory: Supplies the foundational genotypic substrate—canonical driver mutations—that initiate ground state perturbation and directly impair Φ;

2. "Bad luck" theory: Acts as a stochastic accelerator of genetic variation accumulation, increasing the probability of ground state destabilization without altering its fundamental logic;

3. Tissue field theory: Functions as a microenvironmental amplifier—modulating both the rate of Φ erosion and the sensitivity of Ψ activation thresholds—thereby determining whether a given Φ deficit translates into phenotypic progression;

4. Functional Ground State Theory: Acts as the intuitive, dynamic phenotypic carrier of Φ, while also serving as the central transducer—integrating upstream genetic, stochastic, and environmental inputs into coordinated phenotypic reprogramming that directly drives Φ impairment.

In essence, tumorigenesis is not an active program of aggression, but a passive, system-level survival response to progressive Φ decline—rooted in the evolutionary trade-offs inherent to multicellular organization.

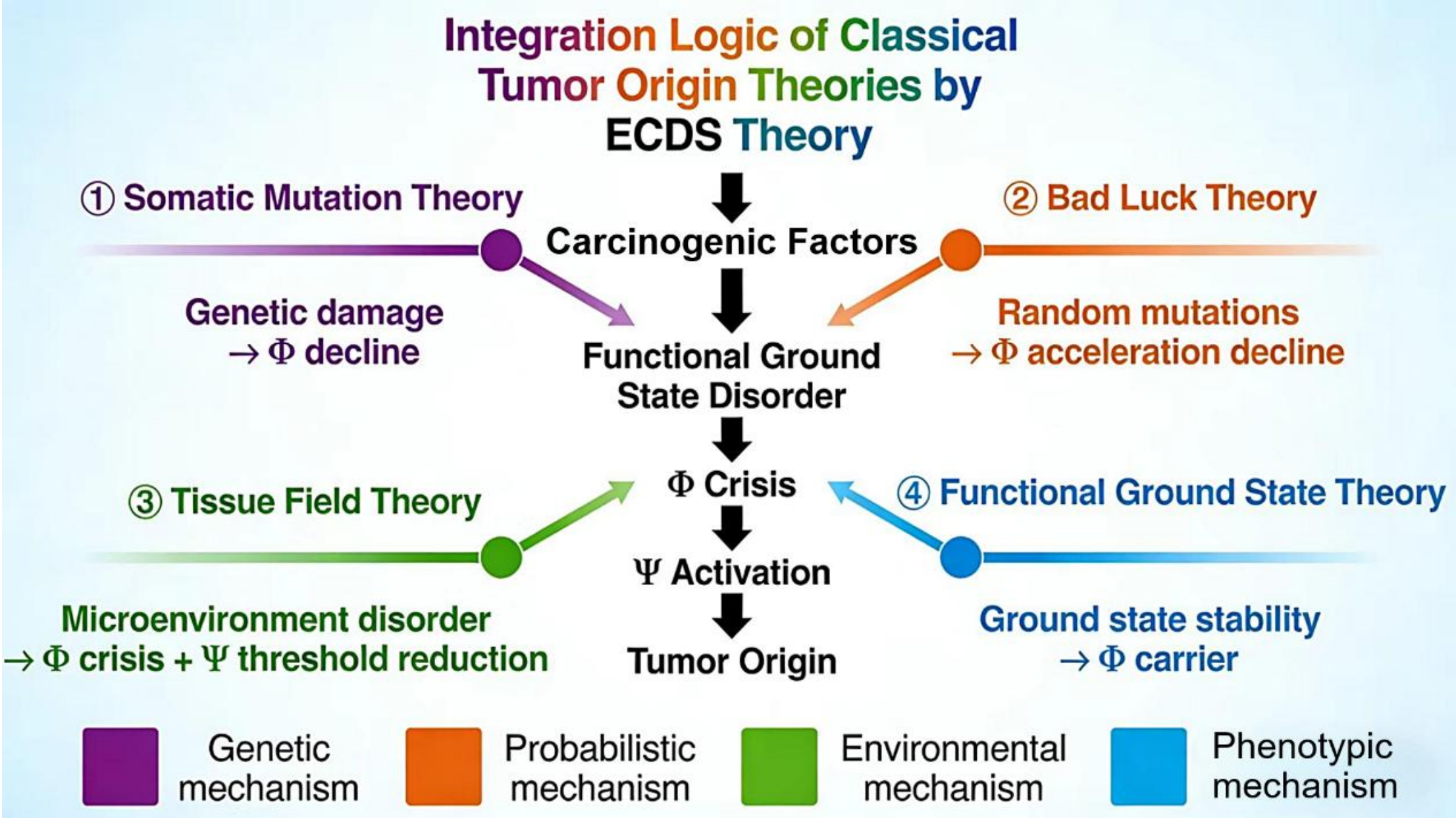

**Figure 3. The ECDS Theory as a unifying framework for classical tumor origin theories**

Guided by driver mutation-initiated Φ crisis → non-genetic amplification → passive Ψ activation, the four theories play complementary roles: genotypic trigger (somatic mutation), stochastic accelerator ("bad luck"), microenvironmental modulator (tissue field), and phenotypic carrier of Φ (functional ground state). Collectively, they drive unidirectional Φ erosion, necessitating Ψ activation to maintain $\Phi + \Psi \geq \Theta$ and triggering tumor initiation.

## 4. Integrative Analysis of Tumor Evolution Stages via the ECDS Framework

Tumor origin—initiated by driver mutation-induced Φ impairment and consequent passive Ψ activation—lays the biological foundation for malignant progression. The dynamic Φ–Ψ coupling continues to evolve through sequentially definable stages, shaped by progressive Φ erosion and hierarchical Ψ escalation (**Figure 4**). Based on quantitative shifts in ECDS core parameters (Φ magnitude, Ψ dimensionality, Θ compliance) and corresponding histopathological phenotypes, tumor evolution is delineated into four non-redundant stages. Cumulative driver mutations, amplified by microenvironmental stressors, deepen Φ impairment; crucially, Ψ upregulation in each tier exacerbates Φ decline, with malignant phenotypes directly reflecting Ψ intensity. This stage-specific pattern embodies the core ECDS axiom (Φ impairment → Ψ activation), forming a continuous trajectory where escalating Φ erosion mandates hierarchical Ψ upgrading and heightened malignancy.

### 4.1 Tumor Initiation Stage: Φ Reaches Oncogenic Threshold; Initial Ψ Activation (Precancerous Lesion / Early In Situ Carcinoma)

Φ falls below the threshold for autonomous persistence, with mild impairment (preserved genomic integrity, subclinical metabolic perturbations, modestly compromised microenvironmental adaptability). Physiological compensation fails to sustain $\Phi + \Psi \geq \Theta$, prompting cells to engage hyperproliferation as the low-complexity survival strategy. Immune evasion, invasion, and stromal remodeling pathways remain silent, establishing a transient metastable Φ–Ψ equilibrium—the earliest molecularly defined oncogenic state.[11,36]

### 4.2 Tumor Progression Stage: Sustained Φ Decline and Ψ Diversification (Advanced In Situ Carcinoma)

Φ erosion intensifies via ongoing mutations and microenvironmental deterioration. Ψ transitions from monofunctional to multifunctional activation (proliferation + anti-apoptotic signaling + metabolic reprogramming + angiogenic induction). This diversification increases cellular dependence on extrinsic nutrients and growth factors, impairs redox homeostasis, and narrows survival scope—reinforcing the self-reinforcing feedback loop: Φ impairment → Ψ diversification → accelerated Φ decline. Hyperproliferation exacerbates replication stress and telomere erosion, reinforcing the Φ-Ψ feedback loop—a core driver of malignant progression.[37]

### 4.3 Tumor Local Invasion Stage: Profound Φ Impairment and Invasion–Metastasis–Focused Ψ (Locally Advanced Carcinoma)

Diversified Ψ-driven clonal expansion exacerbates resource competition and microenvironmental collapse, driving profound Φ impairment. Microenvironmental adaptability fails (coordinated E-cadherin downregulation, PD-L1 upregulation, stromal degradation), prompting invasion–metastasis as the principal survival modality. Complementary Ψ axes (enhanced proliferation, anti-apoptosis, advanced immune evasion, ECM remodeling) synergize to support invasion, clinically manifesting as basement membrane breach, adjacent tissue infiltration, increased intratumoral heterogeneity, and reduced therapeutic responsiveness.[38-40]

### 4.4 Tumor Distant Metastasis Stage: Extreme Φ Impairment and Adaptive Colonization Ψ (Metastatic Carcinoma)

Φ reaches near-catastrophic impairment (collapsed survival stability, severe mitochondrial dysfunction, near- complete extrinsic dependence). Tumor cells activate high-fidelity adaptive programs (EMT for detachment, anoikis resistance for suspension survival, cytoskeletal remodeling for intravasation). Circulating tumor cells engage “adaptive colonization” Ψ (anti-anoikis signaling

+ stromal co-option + ECM remodeling + context-specific metabolic reprogramming) in distant microenvironments. Distant metastasis is a crisis-driven adaptation, clinically presenting as highly heterogeneous metastatic deposits, with therapeutic objectives shifting from curative eradication to durable disease control.[41,42]

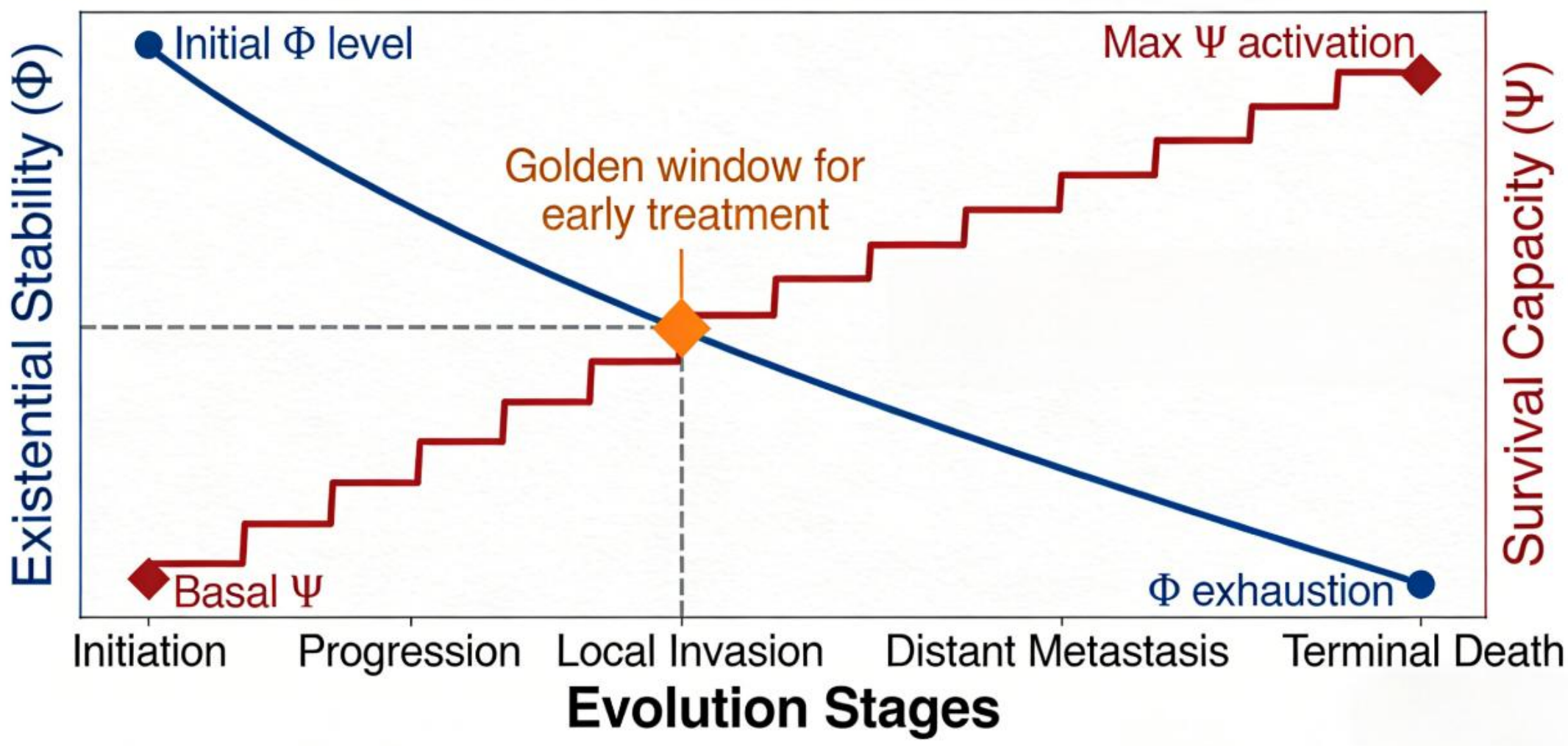

**Figure 4. Dynamic coupling of Φ and Ψ across tumor evolutionary stages**

Φ undergoes continuous, irreversible decline; Ψ responds passively and hierarchically to maintain $\Phi + \Psi \geq \Theta$. The "Golden window for early intervention" lies in the Initiation stage, where Φ impairment is mild and Ψ activation is reversible. Stable Φ–Ψ equilibrium (tumor dormancy) can be achieved at any stage via clinical modulation (e.g., targeted therapy, immunomodulation, microenvironment normalization).

## 5. Reconstruction of Classical Cancer Hallmarks via the ECDS Framework

The fourteen canonical cancer hallmarks encompass the phenotypic spectrum of malignant tumors; while traditional interpretations have sought to offer unified explanations and systematic analyses, they fail to achieve the comprehensive integration embodied by our ECDS framework—instead treating these hallmarks as isolated traits rather than a tightly integrated adaptive network.[6] The ECDS Theory reconceptualizes them as functionally ordered, threshold-governed dimensions of Ψ—sequentially engaged with graded Φ impairment per the principle: graded Φ decline → hierarchical survival pathway activation → maintenance of $\Phi + \Psi \geq \Theta$. Their spatiotemporal hierarchy, cross-level synergy, and focus on cellular persistence transform fragmented traits into a unified, evolution-grounded, stage-predictive framework, mirroring tumor evolutionary trajectories. These hallmark activations are Gradient Activation Model (GAM)-governed phenotypic expressions, with molecular details elaborated in subsequent heterogeneity and resistance analyses. Notably, this framework seamlessly accommodates emerging hallmark-related insights (e.g., phenotypic plasticity, tumor innervation, polymorphic microbiomes) by anchoring them to the core Φ–Ψ coupling logic.

### 5.1 Crisis-Initiation Trigger: Genomic Instability and Mutational Burden—The Primary Driver of Φ Impairment

Genomic instability and mutational accumulation are the foundational trigger for all hallmarks: initiating irreversible Φ erosion and providing the genotypic substrate for Ψ escalation. Driver mutations, structural genomic aberrations (e.g., ecDNA), and DNA repair deficiencies collectively compromise Φ's three core dimensions (survival stability, duration, scope). This trigger operates continuously across stages, intensifies with Φ impairment, and links tumor origin to progression as a unifying mechanistic bridge. Consistent with emerging evidence, non-mutational epigenetic reprogramming—another key initiator of Φ impairment—synergizes with genomic instability to disrupt Φ integrity, laying the groundwork for early Ψ activation (e.g., basal proliferative signaling).[7,43]

### 5.2 Foundational Survival Pathways: Core Mechanisms Enabling Clonal Persistence

When genomic instability drives $\Phi < \Theta$ (without compensatory Ψ), cells engage foundational survival pathways centered on two objectives: sustaining proliferation to offset diminished survival scope/duration, and evading intrinsic/extrinsic elimination. These pathways represent the minimal phenotypic investment for early clonal expansion, with their activation tightly regulated by GAM's low-level activation stage.

#### 5.2.1 Core Proliferative Hallmarks

Sustained proliferative signaling, growth suppressor evasion, and limitless replicative potential synergistically expand cell populations, compensating for compromised Φ. As the basal Ψ tier, they respond to mild-to-moderate Φ impairment, underpinning the initiation-to-progression transition. Hyperproliferation exacerbates replication stress and telomere erosion, reinforcing the Φ impairment → Ψ activation → further Φ erosion feedback loop. Notably, this pathway is dynamically modulated by systemic factors that accelerate Φ erosion and lower Ψ activation thresholds.[3-6]

### 5.2.2 Core Defense Hallmarks

Resistance to cell death (e.g., apoptosis, ferroptosis, necroptosis) and senescence evasion safeguard genomically compromised cells. Immune evasion—via checkpoint ligand upregulation (PD-L1), antigen presentation defects, or immunosuppressive cytokines (TGF-β, IL-10)—establishes "immune sanctuaries" for Φ–Ψ equilibrium. Immune evasion matures incrementally: from early MHC-I downregulation to multi-layered suppression (T-cell exhaustion, Treg recruitment) in advanced stages, a process amplified by tumor-promoting inflammation that lowers Ψ activation thresholds, bridging foundational and adaptive survival strategies. This maturation process is closely coupled to Φ's progressive impairment and TME cell reprogramming (e.g., TAMs-induced immunosuppression).[44]

## 5.3 Supporting Survival Pathways: Ensuring Nutrient and Energetic Homeostasis

Rapid clonal expansion outstrips vascular supply and nutrients, intensifying Φ impairment. Tumor cells activate supporting pathways for metabolic homeostasis, which correspond to GAM's intermediate activation stage and involve dynamic crosstalk with TME components:

- Angiogenic induction: Secretion of pro-angiogenic factors stimulates neovascularization, scaling with Φ impairment but increasing systemic vulnerability via structurally abnormal vessels (poor pericyte coverage).[45,46] Notably, tumor vasculature is not only a nutrient channel but also a modulator of Φ–Ψ coupling—endothelial cells and pericytes secrete paracrine factors that further dampen Φ integrity and enhance Ψ activation;
- Tumor-specific metabolic reprogramming: e.g., the Warburg effect (aerobic glycolysis) enables rapid ATP generation under hypoxia, but induces glucose dependency and disrupts mitochondrial function—compromising Φ's survival stability. This reprogramming is not static: cancer cells exhibit metabolic plasticity (e.g., lactate shuttle with CAFs) to adapt to Φ's deepening impairment.[47]

These intermediate-tier Ψ activations facilitate large-scale proliferation and drive intratumoral heterogeneity via selective pressure.

## 5.4 Adaptive Survival Pathways: Phenotypic and Microenvironmental Plasticity

As Φ deteriorates, the "proliferation–defense–nutrition" triad becomes insufficient, prompting adaptive survival pathways via two strategies: intrinsic phenotypic plasticity and extrinsic microenvironmental exploitation. Both enhance short-term survival at the cost of long-term Φ integrity, driving heterogeneity and therapeutic resistance, and correspond to GAM's high-level activation stage.

### 5.4.1 Self-Adaptation Hallmarks

Unlocking phenotypic plasticity—a core adaptive trait highlighted in recent updates—serves as the central Ψ dimension in this tier. Phenotypic plasticity (dedifferentiation, lineage infidelity, epithelial-mesenchymal plasticity) is enabled by non-mutational epigenetic reprogramming (e.g., histone modification, DNA methylation erasure) and loss of lineage-defining transcription factors (e.g., TP53 dysfunction).[48,49] It destabilizes cellular identity, increases transcriptional/proteomic imbalance, and accelerates Φ decline via compromised genomic and metabolic homeostasis. This plasticity is not stochastic but GAM-governed: as Φ impairment deepens, it transitions from reversible phenotype switching to irreversible epigenetic canalization, laying the foundation for therapy resistance.

### 5.4.2 Microenvironment-Exploitation Hallmarks

Dysbiosis-mediated genotoxicity, immune modulation, and metabolic reprogramming—key contributions of polymorphic microbiomes—buffer existential threats by modulating Φ–Ψ coupling (e.g., gut microbiota-induced inflammation accelerates Φ erosion, while certain microbial metabolites lower Ψ activation thresholds).[50] Concurrently, senescent stromal cells secrete SASP (IL-6, IL-8, MMPs) that stimulates epithelial proliferation, inhibits apoptosis, and reinforces angiogenesis.[51] Immune evasion evolves into a multi-layered defense network—integrating CAFs-mediated ECM remodeling, TAMs-induced immunosuppression, and neuronal signaling that safeguards invasive fronts while deepening Φ impairment.[52] Additionally, tumor innervation (a newly recognized enabling factor) functions as a local modulator: neurotransmitters (e.g., norepinephrine) secreted by intratumoral axonal projections directly enhance Ψ activation (proliferation, invasion) and accelerate Φ erosion via ROS production. These high-order Ψ activations (under severe Φ impairment) are key drivers of heterogeneity and resistance, governed by GAM.

## 5.5 Ultimate Survival Pathway: Invasion and Metastasis—Transcending Spatial Existential Constraints

When foundational, supporting, and adaptive pathways fail to sustain $\Phi + \Psi \geq \Theta$, tumor cells deploy invasion and metastasis as the terminal survival strategy—transcending spatial confinement to access new niches. As the most intricate and costly Ψ program, it abolishes autonomous persistence capacity, maximizes exogenous dependency (e.g., pre-metastatic niche reliance), and minimizes intrinsic survival stability—empirically validating the Principle of Survival Futility. This pathway corresponds to GAM's epigenetic canalization stage, driven by extreme Φ impairment (near-catastrophic genomic/metabolic dysfunction). Notably, metastatic progression is not solely dependent on genetic mutations but also on non-genetic factors: phenotypic plasticity (EMT/MET switching), TME co-option (vascular co-option, neuronal signaling), and systemic cues (obesity-induced metabolic rewiring)—all of which converge on reinforcing Ψ's ultimate activation while accelerating Φ exhaustion. This defines the biological essence of distant metastatic disease.[41]

### 5.6 Cross-Stage Regulatory Axis: Tumor-Promoting Inflammation—A Universal Modulator of Φ–Ψ Coupling

Tumor-promoting inflammation functions as a pervasive, cross-stage regulatory axis and the principal "Φ–Ψ coupling modulator" that accelerates Φ impairment (via oxidative DNA damage, chromosomal instability) and lowers Ψ activation thresholds. It synergizes with other cross-stage factors (tumor innervation, microbiome dysbiosis) to modulate Φ–Ψ dynamics across all GAM stages: in early stages, it primes Φ erosion for initial Ψ activation; in advanced stages, it reinforces adaptive Ψ pathways (plasticity, immune evasion) and facilitates the ultimate transition to invasion and metastasis. This multi-stage modulation aligns with the integrated view of cancer as a systemic disease, linking local hallmark activation to global physiological perturbations.[52-54]

## 6. In-Depth Analysis of Tumor Heterogeneity Based on the ECDS Theory

The hierarchical, threshold-governed activation of cancer hallmarks generates functionally divergent subclonal populations—the defining cellular substrate of tumor heterogeneity.[55] Critically, this phenotypic diversity is not stochastic but is mechanistically governed by the ECDS-specific Gradient Activation Model (GAM), which constitutes the core regulatory architecture for context-dependent survival pathway specification. Within the ECDS framework, tumor heterogeneity is formally defined as: "a hierarchical, functionally redundant, and dynamically regulated network of survival pathways—orchestrated by spatial and temporal gradients of Φ impairment and modulated by developmental constraints, epigenetic plasticity, and spatiotemporal heterogeneity within the tumor microenvironment. (**Table 1**) " This definition unifies the ontogenetic origin, structural organization, and adaptive regulation of heterogeneity—directly embedding it within the foundational ECDS axiom: Φ impairment → Ψ activation.

**Table 1: Unified Correlation of GAM Stages Across Core Chapters in the ECDS Framework**

| GAM Stages | Evolutionary Stage | Hallmark Reconstruction Tier | Heterogeneity Manifestation | Therapeutic Resistance Stage |
|---|---|---|---|---|
| Silent | Initiation Stage | Trigger (Genomic Instability) | No significant heterogeneity | — |
| Low-level activation | Early Progression Stage | Foundational Pathways (Proliferation/Defense) | Mild reversible heterogeneity | Primary Resistance |
| High-level activation | Advanced Progression → Local Invasion | Supporting/Adaptive Pathways (Metabolism/Plasticity) | Significant stable heterogeneity | Secondary Resistance |
| Epigenetic canalization | Distant Metastasis Stage | Ultimate Pathway (Invasion and Metastasis) | Maximized fixed heterogeneity | Malignant Resistance |

Table legend: This table synthesizes the consistent correlations between the four stages of the Gradient Activation Model (GAM) and core oncological processes across the key chapters of the ECDS framework. The "Evolutionary Stage" corresponds to the four-phase pattern of tumor progression (Chapter 4); the "Hallmark Reconstruction Tier" aligns with the hierarchical activation of cancer hallmarks (Chapter 5); the "Heterogeneity Manifestation" reflects the GAM-driven phenotypic diversity of tumors (Chapter 6); and the "Therapeutic Resistance Stage" links to the hierarchical leap mechanism of **Ψ** (Chapter 7). This unified mapping underscores GAM as the central regulatory link that integrates tumor evolution, hallmark acquisition, cancer heterogeneity, and therapeutic resistance within the ECDS framework.

### 6.1 Dynamic Evolution of Tumor Heterogeneity: A Gradient Activation–Driven Mechanism

Tumor heterogeneity arises from the continuous, Φ-gradient–driven activation of latent survival programs—a process formalized as the Gradient Activation Model (GAM). GAM describes a deterministic, stepwise transition of survival pathways through four epigenetically defined functional states in response to escalating Φ damage: silence → low-level activation → high-level activation → epigenetic canalization. Each transition reflects an adaptive recalibration of the Φ–Ψ equilibrium and culminates in the emergence of phenotypically distinct, functionally specialized subclones.

This GAM process is largely contingent upon the magnitude and spatial distribution of Φ impairment—and thus exhibits close alignment with both the severity of existential crisis and the evolutionary stage of the tumor:

- Silent stage (mild Φ impairment): Survival pathways generally remain transcriptionally repressed and epigenetically silenced—mediated by repressive histone marks (e.g., H3K27me3) and promoter DNA methylation—representing the latent reservoir of adaptive potential. This state corresponds to the tumor initiation stage (pre-malignant lesions or dormancy).

- Low-level activation stage (moderate Φ impairment): Subthreshold Φ erosion triggers selective, reversible engagement of basal survival modules—primarily hyperproliferative signaling and anti-apoptotic activity. These transient adaptations are epigenetically labile and phenotypically plastic, contributing modest, reversible heterogeneity; this state aligns with the early tumor progression stage.[56]

- High-level activation stage (severe Φ impairment): Sustained Φ erosion recruits complementary, functionally synergistic survival axes—including metabolic reprogramming, immune evasion, and angiogenic competence—integrating them into a robust, multi-layered, and functionally redundant pathway network. Heterogeneity expands markedly, with subclones acquiring stable, co-adapted phenotypes capable of coordinated resource acquisition and stress resistance. This state correlates with the advanced progression to local invasion stage.

- Epigenetic canalization stage (extreme Φ impairment): Persistent, near-catastrophic Φ damage drives large-scale chromatin remodeling—particularly enhancer stabilization (e.g., super-enhancer formation at key oncogene loci) and promoter demethylation—locking activated survival pathways into constitutively expressed, heritable states. This stage generates phenotypically fixed, therapy-refractory subclones whose malignant traits are maintained independently of ongoing environmental cues. Epigenetic canalization is mediated by coordinated histone modifications and locus-specific DNA hypomethylation[55]. This state dominates the distant metastasis stage.[57]

Heterogeneity Amplification Cycle: External stressors—including progressive mutational accumulation, chronic inflammation, hypoxia, and therapeutic pressure—exacerbate Φ impairment, thereby propelling survival pathways toward higher-order GAM stages. The resulting phenotypic adaptations further degrade microenvironmental fitness and accelerate Φ erosion. This establishes a self-reinforcing feedback loop: Φ impairment drives GAM stage advancement → adaptive phenotypic remodeling degrades microenvironmental fitness → amplified Φ impairment recruits broader survival pathways → further expanding intratumoral heterogeneity. This cycle drives rapid, non-linear expansion of intratumoral heterogeneity while rigorously reinforcing the core ECDS logic: impairment drives adaptation, which in turn exacerbates further impairment.[58]

Tumor evolutionary stages are strictly synchronized with GAM dynamics: The initiation stage (mild Φ impairment) aligns with the GAM silent phase, during which only a small number of cells weakly activate basal proliferative pathways, with no significant intratumoral heterogeneity; the progression stage (moderate Φ impairment) is marked by heterogeneous Φ damage across the lesion, leading to the co-occurrence of low-level and high-level GAM activation, alongside spatially structured expansion of intratumoral heterogeneity; the invasion and metastasis stages (severe to extreme Φ impairment) are dominated by the GAM epigenetic canalization phase in metastatic subclones, which is characterized by maximal functional diversification, phenotypic stability, and intrinsic therapeutic resistance.

This hierarchical, gradient-driven differentiation of survival pathways—governed by the Gradient Activation Model (GAM)—is the principal mechanism sustaining the ECDS viability constraint ($\Phi + \Psi \geq \Theta$) across evolving clonal architectures. It therefore constitutes the mechanistic cornerstone of tumor evolutionary progression and therapeutic failure.

## 7. Core Mechanisms of Therapeutic Resistance in ECDS: A Hierarchical Leap Perspective

Classical models of therapeutic resistance emphasize clonal selection of pre-existing, genetically encoded advantageous variants. In contrast, the ECDS Theory proposes a mechanistically distinct, evolutionarily grounded framework: therapeutic intervention does not merely select resistant clones but actively exacerbates Existential Stability (Φ) impairment. This imposes selective pressure that drives tumor cells to execute hierarchical, non-stochastic "Ψ leaps" across Gradient Activation Model (GAM) stages—from low-level activation toward epigenetic solidification—governed by the dual principles of adaptive Ψ upregulation and Survival Futility. Critically, the Φ–Ψ coupling mechanism embeds resistance within the core ECDS logic (Φ impairment → Ψ activation), unifying resistance phenomena across diverse therapeutic modalities (cytotoxic chemotherapy, targeted therapy, immunotherapy, and radiotherapy) under a single predictive theoretical scaffold.[59,60]

### 7.1 Core Mechanism: Hierarchical Ψ Leap

Therapeutic resistance originates from hierarchical Ψ leaps rooted in GAM-governed intratumoral heterogeneity: subclones in different GAM stages exhibit heterogeneous responses to therapy, which amplifies selective pressure for Ψ escalation. These leaps reflect a structured, stage-specific "hierarchical upgrade" of Survival Capacity (Ψ), directly driven by therapy-induced Φ impairment—consistent with ECDS core axioms. Since tumor cells have already surpassed physiological compensation thresholds, each Ψ leap correlates with worsening Φ damage, increased malignancy, and reduced reversibility, adhering to the GAM-defined hierarchy of cancer hallmarks (**Figure 5**).

- Primary leap (basic resistance): Low-level activation of latent pathways engages foundational hallmarks (sustained proliferation, apoptosis evasion). Inhibition of a single pathway (e.g., EGFR signaling) triggers pre-existing genetic backup alternatives (e.g., MET amplification)—a "backup circuit" maintained by genetic alterations that remains pharmacologically tractable via targeted inhibition of the compensatory node.[61,62]

- Secondary leap (moderate resistance): High-level activation recruits supportive/adaptive hallmarks (angiogenesis, metabolic reprogramming, phenotypic plasticity). Prolonged therapeutic pressure exceeds single-pathway compensation, forcing coordinated engagement of lineage infidelity, OXPHOS-to-glycolysis shifts, and stromal co-option—resulting in heterogeneous resistance that necessitates rationally designed combination therapies.[63]

- Tertiary leap (malignant resistance): Epigenetic canalization activates ultimate hallmarks (invasion, metastasis, immune sanctuary formation). Excessive therapeutic aggression (e.g., high-dose cytotoxic therapy) induces genotoxic stress and metabolic dysfunction, driving near-catastrophic Φ impairment, with stable chromatin remodeling locking survival pathways into heritable, constitutively active states[56]. Clinically manifesting as rapid disease progression and refractoriness to salvage therapies, this leap is preventable via kinetic modulation of Φ impairment (e.g., dose de-escalation, intermittent scheduling, or concurrent cytoprotective agents).[62]

The hierarchical Ψ leap evolves sequentially from "redundant pathway compensation" (Primary) to "adaptive phenotypic reconstruction" (Secondary) and "spatial niche escape" (Tertiary), forming a self-reinforcing cycle: therapy → Φ exacerbation → Ψ leap → further Φ erosion → intensified therapeutic pressure.

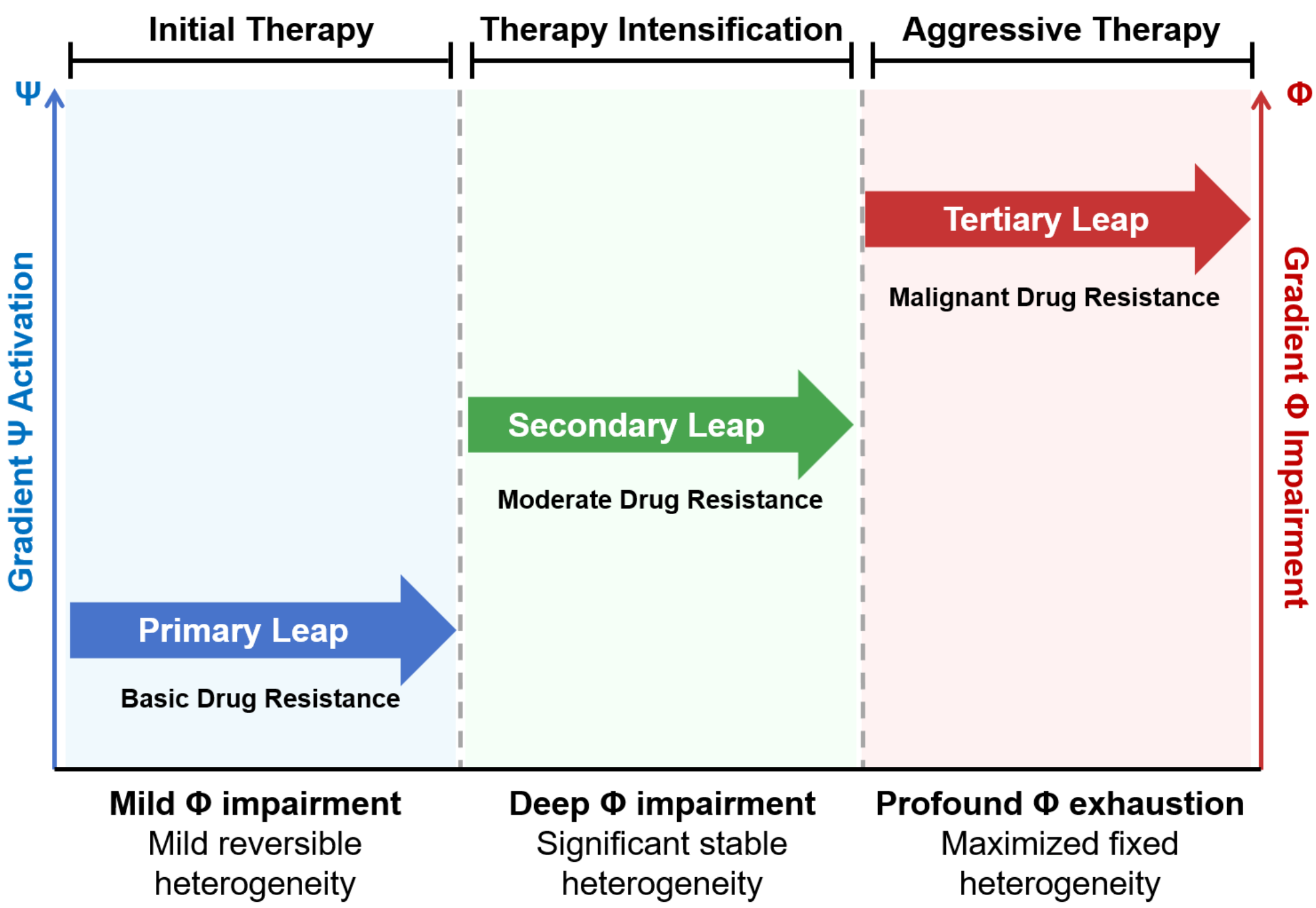

**Figure 5. Schematic Illustration of GAM-Driven Gradient Heterogeneity and Hierarchical Ψ Leaps Mediating Therapeutic Resistance**

This figure illustrates the core mechanism underlying therapeutic resistance, which arises from the interplay between Gradient Activation Model (GAM)-driven gradient heterogeneity and hierarchical Ψ leaps within the ECDS framework. Therapeutic intervention exacerbates Existential Stability (Φ) impairment in a dose- and kinetics-dependent manner, acting on pre-existing intratumoral heterogeneity defined by subclones distributed across distinct GAM stages. This selective pressure induces non-stochastic, hierarchical "Ψ leaps" across GAM-aligned functional stages: (1) low-level activation of foundational survival pathways, corresponding to mild reversible heterogeneity and primary resistance; (2) high-level activation of supportive/adaptive survival pathways, associated with significant stable heterogeneity and secondary resistance; and (3) epigenetic canalization of ultimate survival pathways, linked to maximized fixed heterogeneity and malignant resistance. Each leap correlates with progressive Φ damage, heightened malignancy, and reduced reversibility, where GAM-driven gradient heterogeneity amplifies the adaptive potential for Ψ escalation.

### 7.2 ECDS–Based Therapeutic Resistance Intervention Strategies

Guided by the hierarchical Ψ leap mechanism, the ECDS Theory proposes three synergistic strategies to intercept resistance at its source, shifting the paradigm from reactive "pathway blockade" to proactive "Φ–Ψ equilibrium regulation."

- Rate Control of Φ Impairment: Avoid supraphysiological or uncontrolled Φ erosion (e.g., the type that triggers epigenetic canalization). Implement "precision impairment modulation" (e.g., targeted agents plus low-dose immunomodulators or cytoprotective antioxidants) to keep Φ within a therapeutic window, confining Ψ to reversible activation tiers for durable disease control.[64]
- Latent Program Inhibition: Target epigenetic machinery (HDACi, BET antagonists, EZH2 inhibitors) that sustains transcriptionally poised survival pathways. Reducing this "backup" pool constrains Ψ adaptability, prevents GAM stage progression, and synergizes with standard therapies to eliminate pre-existing or therapy-inducible resistance.[65]
- Microenvironmental Stabilization: Modulate the tumor microenvironment as a Φ-buffering system (e.g., anti-inflammatory agents, vascular normalization, stromal reprogramming). These interventions raise Ψ activation thresholds, dampen Φ erosion kinetics, and restore therapeutic sensitivity in resistant subclones by reinforcing Φ–Ψ homeostasis.[66,67]

Collectively, these axes (Φ impairment kinetics, latent program reserve, microenvironmental stability) constitute actionable clinical paradigms, with implementation details elaborated in subsequent sections on clinical implications.

## 8. Clinical Implications and Future Perspectives of the ECDS Theory

### 8.1 Paradigm Shift in Clinical Oncology

The ECDS Theory reshapes clinical cancer cognition: tumor cells are not intrinsically aggressive, but Φ erosion–driven passive survival systems—catalyzing a paradigm shift from "confrontational eradication" to "systemic equilibrium regulation."[68] Therapeutic goals thus shift from maximal cytoreduction (which risks catastrophic Φ impairment and irreversible Ψ leap to epigenetic canalization) to long-term maintenance of Φ–Ψ balance, aligning with chronic disease management principles for advanced malignancies.

For early-stage tumors, the "golden window" lies in reversing mild pre-malignant Φ impairment (e.g., enhancing DNA repair or redox homeostasis) before latent survival pathways enter low-level activation. For advanced tumors, the priority is modulating the intensity of Ψ activation to prevent progression to epigenetic canalization—mitigating malignancy, delaying metastasis, and addressing therapy-induced metastatic acceleration.[69]

To illustrate this clinical stratification and therapeutic personalization based on ECDS principles, we take advanced non-small cell lung cancer (NSCLC) as an example: Φ gradient markers (e.g., composite scores of mutational burden, mitochondrial dysfunction, and stromal immune exclusion) stratify GAM stages: high-level activation (e.g., co-expression of glycolytic enzymes, PD-L1, VEGF) warrants “Rate Control of Φ Impairment + Microenvironmental Stabilization”; epigenetic canalization (e.g., H3K27ac enrichment at EMT loci) requires “Latent Program Inhibition + Targeted Ψ Pathway Blockade” to constrain progression.[70,71]

Three complementary biomarker classes enable precision intervention: (1) Φ impairment gradient markers: Integrate multi-dimensional readouts (genomic instability, metabolic dysregulation, microenvironmental maladaptation);[72] (2) Ψ activation stage markers: Stage-specific transcriptional/epigenetic/proteomic signatures (low-level: ERK phosphorylation; high-level: LDHA/HK2 co-expression; epigenetic canalization: super-enhancer occupancy at SNAI1);[73] (3) Latent program markers: Quantify epigenetic plasticity (chromatin accessibility via ATAC-seq, histone modification ratios, epigenetic regulator expression).

These biomarkers form a predictive cascade (Φ impairment → GAM stage → resistance risk) that supports dynamic, individualized decisions—shifting oncology from static histopathology-driven algorithms to real-time system-level adaptive management.

### 8.2 Future Research Directions

While the ECDS Theory boasts robust integrative explanatory power, it requires rigorous experimental validation and methodological refinement. To address unresolved questions (e.g., quantitative measurement of Φ-Ψ equilibrium, GAM transition kinetics), three interdependent research priorities are proposed to advance quantification, mechanistic understanding, and clinical translation:

- Quantitative Φ–Ψ–Θ Dynamics: Develop and prospectively validate a standardized multi-dimensional scoring system for Φ (encompassing genomic, metabolic, and microenvironmental metrics) and Ψ (including pathway activation intensity, functional redundancy, and GAM stage enrichment). This framework should enable precise longitudinal assessment of the $\Phi + \Psi \geq \Theta$ threshold, laying a quantitative foundation for individualized balance-oriented interventions.[74]

- GAM Dynamics Validation: Leverage PDX models, 3D organoids, and longitudinal single-cell multi-omics to characterize spatiotemporal transitions of survival pathways across GAM stages under therapeutic stress. Key objectives include: mapping kinetic thresholds for stage transitions; elucidating epigenetic mechanisms that regulate pathway activation and reversibility; and quantifying how microenvironmental factors modulate transition probability. This work will validate the core hypothesis (GAM-driven heterogeneity and resistance) while informing Φ–Ψ quantification and stage-targeted regimens.[75,76]

- Clinical Translation of Balance-Regulating Strategies: Launch well-designed phase I/II trials to evaluate ECDS-guided combination regimens, with assessments of safety, pharmacodynamic effects on Φ–Ψ equilibrium, preliminary efficacy, and dose–response relationships.

### 8.4 Conclusion

The ECDS Theory establishes a unified, evolutionarily grounded meta-theoretical framework for oncology—synthesizing historically fragmented classical theories through the singular, causally coherent logic of Φ impairment–driven Ψ activation. It reinterprets tumor origin, multistage evolution, hallmark acquisition, intratumoral heterogeneity, and therapeutic resistance not as isolated phenomena, but as interconnected, stage-dependent manifestations of a single, survival-compensatory process. Critically, this framework reveals the intrinsic vulnerability of malignancy: its absolute, non-redundant dependence on Ψ-mediated adaptation to sustain viability amid progressive, irreversible Φ erosion. By shifting the therapeutic imperative from "eradicating malignant phenotypes" to "dynamically regulating Φ–Ψ equilibrium and GAM progression," the ECDS Theory provides a transformative conceptual scaffold for both fundamental cancer research and precision clinical oncology—paving the way for a new era of system-informed, adaptive, and patient-centered cancer management.

## References

1. Collaborators, G.B.D.C. (2025). The global, regional, and national burden of cancer, 1990-2023, with forecasts to 2050: a systematic analysis for the Global Burden of Disease Study 2023. Lancet *406*, 1565-1586. 10.1016/S0140-6736(25)01635-6.
2. Siegel, R.L., Kratzer, T.B., Giaquinto, A.N., Sung, H., and Jemal, A. (2025). Cancer statistics, 2025. CA Cancer J Clin *75*, 10-45. 10.3322/caac.21871.
3. Hanahan, D., and Weinberg, R.A. (2000). The hallmarks of cancer. Cell *100*, 57-70. 10.1016/s0092-8674(00)81683-9.
4. Hanahan, D. (2022). Hallmarks of Cancer: New Dimensions. Cancer Discov *12*, 31-46. 10.1158/2159-8290.CD-21-1059.
5. Hanahan, D., and Weinberg, R.A. (2011). Hallmarks of cancer: the next generation. Cell *144*, 646-674. 10.1016/j.cell.2011.02.013.
6. Hanahan, D. (2026). Hallmarks of cancer-Then and now, and beyond. Cell.

10.1016/j.cell.2025.12.049.
7. Vogelstein, B., Papadopoulos, N., Velculescu, V.E., Zhou, S., Diaz, L.A., Jr., and Kinzler, K.W. (2013). Cancer genome landscapes. Science *339*, 1546-1558. 10.1126/science.1235122.
8. Jassim, A., Rahrmann, E.P., Simons, B.D., and Gilbertson, R.J. (2023). Cancers make their own luck: theories of cancer origins. Nat Rev Cancer *23*, 710-724. 10.1038/s41568-023-00602-5.
9. Greaves, M., and Maley, C.C. (2012). Clonal evolution in cancer. Nature *481*, 306-313. 10.1038/nature10762.
10. Turajlic, S., Sottoriva, A., Graham, T., and Swanton, C. (2019). Resolving genetic heterogeneity in cancer. Nat Rev Genet *20*, 404-416. 10.1038/s41576-019-0114-6.
11. de Visser, K.E., and Joyce, J.A. (2023). The evolving tumor microenvironment: From cancer initiation to metastatic outgrowth. Cancer Cell *41*, 374-403. 10.1016/j.ccell.2023.02.016.
12. Laplane, L., and Maley, C.C. (2024). The evolutionary theory of cancer: challenges and potential solutions. Nat Rev Cancer *24*, 718-733. 10.1038/s41568-024-00734-2.
13. Lu, H., Yang, M., and Zhou, Q. (2023). Reprogramming transcription after DNA damage: recognition, response, repair, and restart. Trends Cell Biol *33*, 682-694. 10.1016/j.tcb.2022.11.010.
14. Monzel, A.S., Enriquez, J.A., and Picard, M. (2023). Multifaceted mitochondria: moving mitochondrial science beyond function and dysfunction. Nat Metab *5*, 546-562. 10.1038/s42255-023-00783-1.
15. Rossiello, F., Jurk, D., Passos, J.F., and d'Adda di Fagagna, F. (2022). Telomere dysfunction in ageing and age-related diseases. Nat Cell Biol *24*, 135-147. 10.1038/s41556-022-00842-x.
16. Feng, T., Xie, F., Lee, L.M.Y., Lin, Z., Tu, Y., Lyu, Y., Yu, P., Wu, J., Chen, B., Zhang, G., et al. (2025). Cellular senescence in cancer: from mechanism paradoxes to precision therapeutics. Mol Cancer *24*, 213. 10.1186/s12943-025-02419-2.
17. De Martino, M., Rathmell, J.C., Galluzzi, L., and Vanpouille-Box, C. (2024). Cancer cell metabolism and antitumour immunity. Nat Rev Immunol *24*, 654-669. 10.1038/s41577-024-01026-4.
18. Galluzzi, L., Yamazaki, T., and Kroemer, G. (2018). Linking cellular stress responses to systemic homeostasis. Nat Rev Mol Cell Biol *19*, 731-745. 10.1038/s41580-018-0068-0.
19. Klapp, V., Alvarez-Abril, B., Leuzzi, G., Kroemer, G., Ciccia, A., and Galluzzi, L. (2023). The DNA Damage Response and Inflammation in Cancer. Cancer Discov *13*, 1521-1545. 10.1158/2159-8290.CD-22-1220.
20. Liu, L., Michowski, W., Kolodziejczyk, A., and Sicinski, P. (2019). The cell cycle in stem cell proliferation, pluripotency and differentiation. Nat Cell Biol *21*, 1060-1067. 10.1038/s41556-019-0384-4.
21. D.Y., W. (2020). A Unified Theory of Evolution: Natural, Mental and Social (1st ed.) (Bridgeminds).
22. Aktipis, C.A., Boddy, A.M., Jansen, G., Hibner, U., Hochberg, M.E., Maley, C.C., and Wilkinson, G.S. (2015). Cancer across the tree of life: cooperation and cheating in multicellularity. Philos Trans R Soc Lond B Biol Sci *370*. 10.1098/rstb.2014.0219.
23. Greten, F.R., and Grivennikov, S.I. (2019). Inflammation and Cancer: Triggers, Mechanisms, and Consequences. Immunity *51*, 27-41. 10.1016/j.immuni.2019.06.025.
24. Ngaha, T.Y.S., Zhilenkova, A.V., Essogmo, F.E., Uchendu, I.K., Abah, M.O., Fossa, L.T., Sangadzhieva, Z.D., V, D.S., A, S.R., Y, N.P., et al. (2023). Angiogenesis in Lung Cancer: Understanding the Roles of Growth Factors. Cancers (Basel) *15*. 10.3390/cancers15184648.
25. Zhang, C.X., Huang, R.Y., Sheng, G., and Thiery, J.P. (2025). Epithelial-mesenchymal transition. Cell *188*, 5436-5486. 10.1016/j.cell.2025.08.033.
26. Lambert, A.W., Pattabiraman, D.R., and Weinberg, R.A. (2017). Emerging Biological Principles of Metastasis. Cell *168*, 670-691. 10.1016/j.cell.2016.11.037.
27. Baracos, V.E., Martin, L., Korc, M., Guttridge, D.C., and Fearon, K.C.H. (2018). Cancer-associated cachexia. Nat Rev Dis Primers *4*, 17105. 10.1038/nrdp.2017.105.
28. Argiles, J.M., Lopez-Soriano, F.J., Stemmler, B., and Busquets, S. (2023). Cancer-associated cachexia - understanding the tumour macroenvironment and microenvironment to improve management. Nat Rev Clin Oncol *20*, 250-264. 10.1038/s41571-023-00734-5.
29. Risson, E., Nobre, A.R., Maguer-Satta, V., and Aguirre-Ghiso, J.A. (2020). The current paradigm and challenges ahead for the dormancy of disseminated tumor cells. Nat Cancer *1*, 672-680. 10.1038/s43018-020-0088-5.
30. Havel, J.J., Chowell, D., and Chan, T.A. (2019). The evolving landscape of biomarkers for checkpoint inhibitor immunotherapy. Nat Rev Cancer *19*, 133-150. 10.1038/s41568-019-0116-x.
31. Martincorena, I., Roshan, A., Gerstung, M., Ellis, P., Van Loo, P., McLaren, S., Wedge, D.C., Fullam, A., Alexandrov, L.B., Tubio, J.M., et al. (2015). Tumor evolution. High burden and pervasive positive selection of somatic mutations in normal human skin. Science *348*, 880-886. 10.1126/science.aaa6806.
32. Lee-Six, H., Olafsson, S., Ellis, P., Osborne, R.J., Sanders, M.A., Moore, L., Georgakopoulos, N., Torrente, F., Noorani, A., Goddard, M., et al. (2019). The landscape of somatic mutation in normal colorectal epithelial cells. Nature *574*, 532-537. 10.1038/s41586-019-1672-7.
33. Liang, X., Weng, J., You, Z., Wang, Y., Wen, J., Xia, Z., Huang, S., Luo, P., and Cheng, Q. (2025). Oxidative stress in cancer: from tumor and microenvironment remodeling to therapeutic frontiers. Mol Cancer *24*, 219. 10.1186/s12943-025-02375-x.
34. Zhu, L., Finkelstein, D., Gao, C., Shi, L., Wang, Y., Lopez-Terrada, D., Wang, K., Utley, S., Pounds, S., Neale, G., et al. (2016). Multi-organ Mapping of Cancer Risk. Cell *166*, 1132-1146 e1137. 10.1016/j.cell.2016.07.045.
35. Allis, C.D., and Jenuwein, T. (2016). The molecular hallmarks of epigenetic control. Nat Rev Genet *17*, 487-500. 10.1038/nrg.2016.59.
36. Galassi, C., Chan, T.A., Vitale, I., and Galluzzi, L. (2024). The hallmarks of cancer immune evasion. Cancer Cell *42*, 1825-1863. 10.1016/j.ccell.2024.09.010.
37. Gao, Y., Siyu, Z., Zhang, X., Du, Y., Ni, T., and Hao, S. (2024). Crosstalk between metabolic and epigenetic modifications during cell carcinogenesis. iScience *27*,

111359. 10.1016/j.isci.2024.111359.
38. Thiery, J.P., Acloque, H., Huang, R.Y., and Nieto, M.A. (2009). Epithelial-mesenchymal transitions in development and disease. Cell *139*, 871-890. 10.1016/j.cell.2009.11.007.
39. O'Donnell, J.S., Teng, M.W.L., and Smyth, M.J. (2019). Cancer immunoediting and resistance to T cell-based immunotherapy. Nat Rev Clin Oncol *16*, 151-167. 10.1038/s41571-018-0142-8.
40. Bakir, B., Chiarella, A.M., Pitarresi, J.R., and Rustgi, A.K. (2020). EMT, MET, Plasticity, and Tumor Metastasis. Trends Cell Biol *30*, 764-776. 10.1016/j.tcb.2020.07.003.
41. Gerstberger, S., Jiang, Q., and Ganesh, K. (2023). Metastasis. Cell *186*, 1564-1579. 10.1016/j.cell.2023.03.003.
42. Shi, X., Wang, X., Yao, W., Shi, D., Shao, X., Lu, Z., Chai, Y., Song, J., Tang, W., and Wang, X. (2024). Mechanism insights and therapeutic intervention of tumor metastasis: latest developments and perspectives. Signal Transduct Target Ther *9*, 192. 10.1038/s41392-024-01885-2.
43. Martinez-Jimenez, F., Muinos, F., Sentis, I., Deu-Pons, J., Reyes-Salazar, I., Arnedo-Pac, C., Mularoni, L., Pich, O., Bonet, J., Kranas, H., et al. (2020). A compendium of mutational cancer driver genes. Nat Rev Cancer *20*, 555-572. 10.1038/s41568-020-0290-x.
44. Tufail, M., Jiang, C.H., and Li, N. (2025). Immune evasion in cancer: mechanisms and cutting-edge therapeutic approaches. Signal Transduct Target Ther *10*, 227. 10.1038/s41392-025-02280-1.
45. Kabir, A.U., Subramanian, M., Kwon, Y., and Choi, K. (2026). Linking tumour angiogenesis and tumour immunity. Nat Rev Immunol *26*, 35-51. 10.1038/s41577-025-01211-z.
46. Guelfi, S., Hodivala-Dilke, K., and Bergers, G. (2024). Targeting the tumour vasculature: from vessel destruction to promotion. Nat Rev Cancer *24*, 655-675. 10.1038/s41568-024-00736-0.
47. Koppenol, W.H., Bounds, P.L., and Dang, C.V. (2011). Otto Warburg's contributions to current concepts of cancer metabolism. Nat Rev Cancer *11*, 325-337. 10.1038/nrc3038.
48. Nacev, B.A., Feng, L., Bagert, J.D., Lemiesz, A.E., Gao, J., Soshnev, A.A., Kundra, R., Schultz, N., Muir, T.W., and Allis, C.D. (2019). The expanding landscape of 'oncohistone' mutations in human cancers. Nature *567*, 473-478. 10.1038/s41586-019-1038-1.
49. Dongre, A., and Weinberg, R.A. (2019). New insights into the mechanisms of epithelial-mesenchymal transition and implications for cancer. Nat Rev Mol Cell Biol *20*, 69-84. 10.1038/s41580-018-0080-4.
50. de Vos, W.M., Tilg, H., Van Hul, M., and Cani, P.D. (2022). Gut microbiome and health: mechanistic insights. Gut *71*, 1020-1032. 10.1136/gutjnl-2021-326789.
51. Wang, B., Han, J., Elisseeff, J.H., and Demaria, M. (2024). The senescence-associated secretory phenotype and its physiological and pathological implications. Nat Rev Mol Cell Biol *25*, 958-978. 10.1038/s41580-024-00727-x.
52. Roerden, M., and Spranger, S. (2025). Cancer immune evasion, immunoediting and intratumour heterogeneity. Nat Rev Immunol *25*, 353-369. 10.1038/s41577-024-01111-8.

53. Zhao, H., Wu, L., Yan, G., Chen, Y., Zhou, M., Wu, Y., and Li, Y. (2021). Inflammation and tumor progression: signaling pathways and targeted intervention. Signal Transduct Target Ther *6*, 263. 10.1038/s41392-021-00658-5.
54. Martinez-Reyes, I., and Chandel, N.S. (2021). Cancer metabolism: looking forward. Nat Rev Cancer *21*, 669-680. 10.1038/s41568-021-00378-6.
55. Dagogo-Jack, I., and Shaw, A.T. (2018). Tumour heterogeneity and resistance to cancer therapies. Nat Rev Clin Oncol *15*, 81-94. 10.1038/nrclinonc.2017.166.
56. Feinberg, A.P., Koldobskiy, M.A., and Gondor, A. (2016). Epigenetic modulators, modifiers and mediators in cancer aetiology and progression. Nat Rev Genet *17*, 284-299. 10.1038/nrg.2016.13.
57. Fan, Q., Chen, H., Liao, Q., Zhang, C., Ding, Y., Zou, X., and Fu, G. (2023). Distribution of organophosphate ester fractions in sediment of the Eastern China Marginal Seas and the influencing factors. Environ Sci Pollut Res Int *30*, 23810-23819. 10.1007/s11356-022-23921-w.
58. Perez-Gonzalez, A., Bevant, K., and Blanpain, C. (2023). Cancer cell plasticity during tumor progression, metastasis and response to therapy. Nat Cancer *4*, 1063-1082. 10.1038/s43018-023-00595-y.
59. Na, F., Pan, X., Chen, J., Chen, X., Wang, M., Chi, P., You, L., Zhang, L., Zhong, A., Zhao, L., et al. (2022). KMT2C deficiency promotes small cell lung cancer metastasis through DNMT3A-mediated epigenetic reprogramming. Nat Cancer *3*, 753-767. 10.1038/s43018-022-00361-6.
60. Tian, C., Zhou, J., Li, X., Gao, Y., Wen, Q., Kang, X., Wang, N., Yao, Y., Jiang, J., Song, G., et al. (2023). Impaired histone inheritance promotes tumor progression. Nat Commun *14*, 3429. 10.1038/s41467-023-39185-y.
61. Passaro, A., Janne, P.A., Mok, T., and Peters, S. (2021). Overcoming therapy resistance in EGFR-mutant lung cancer. Nat Cancer *2*, 377-391. 10.1038/s43018-021-00195-8.
62. Ge, M., Chen, X.Y., Huang, P., Fleishman, J.S., Yang, D.H., Wu, Z.X., Ke, Z.F., and Chen, Z.S. (2025). Understanding and overcoming multidrug resistance in cancer. Nat Rev Clin Oncol *22*, 760-780. 10.1038/s41571-025-01059-1.
63. Soragni, A., Knudsen, E.S., O'Connor, T.N., Tognon, C.E., Tyner, J.W., Gini, B., Kim, D., Bivona, T.G., Zang, X., Witkiewicz, A.K., et al. (2025). Acquired resistance in cancer: towards targeted therapeutic strategies. Nat Rev Cancer *25*, 613-633. 10.1038/s41568-025-00824-9.
64. Waldman, A.D., Fritz, J.M., and Lenardo, M.J. (2020). A guide to cancer immunotherapy: from T cell basic science to clinical practice. Nat Rev Immunol *20*, 651-668. 10.1038/s41577-020-0306-5.
65. Bates, S.E. (2020). Epigenetic Therapies for Cancer. N Engl J Med *383*, 650-663. 10.1056/NEJMra1805035.
66. Bejarano, L., Jordao, M.J.C., and Joyce, J.A. (2021). Therapeutic Targeting of the Tumor Microenvironment. Cancer Discov *11*, 933-959. 10.1158/2159-8290.CD-20-1808.
67. De Palma, M., Biziato, D., and Petrova, T.V. (2017). Microenvironmental regulation of tumour angiogenesis. Nat Rev Cancer *17*, 457-474. 10.1038/nrc.2017.51.
68. Palmer, A.C., and Sorger, P.K. (2017). Combination Cancer Therapy Can Confer Benefit via Patient-to-Patient Variability without Drug Additivity or Synergy. Cell

*171*, 1678-1691 e1613. 10.1016/j.cell.2017.11.009.
69. Yao, T.H., Treekitkarnmongkol, W., Putluri, N., Sankaran, D., Nguyen, T., Balasenthil, S., Hurd, M.W., Chen, M., Brand, R.E., Lampe, P.D., et al. (2026). Machine learning-based multimodal biomarkers enable accurate diagnosis and early detection of pancreatic ductal adenocarcinoma. Sci Rep *16*, 658. 10.1038/s41598-025-29548-4.
70. Cascone, T., McKenzie, J.A., Mbofung, R.M., Punt, S., Wang, Z., Xu, C., Williams, L.J., Wang, Z., Bristow, C.A., Carugo, A., et al. (2018). Increased Tumor Glycolysis Characterizes Immune Resistance to Adoptive T Cell Therapy. Cell Metab *27*, 977-987 e974. 10.1016/j.cmet.2018.02.024.
71. Hellmann, M.D., Paz-Ares, L., Bernabe Caro, R., Zurawski, B., Kim, S.W., Carcereny Costa, E., Park, K., Alexandru, A., Lupinacci, L., de la Mora Jimenez, E., et al. (2019). Nivolumab plus Ipilimumab in Advanced Non-Small-Cell Lung Cancer. N Engl J Med *381*, 2020-2031. 10.1056/NEJMoa1910231.
72. Litchfield, K., Reading, J.L., Puttick, C., Thakkar, K., Abbosh, C., Bentham, R., Watkins, T.B.K., Rosenthal, R., Biswas, D., Rowan, A., et al. (2021). Meta-analysis of tumor- and T cell-intrinsic mechanisms of sensitization to checkpoint inhibition. Cell *184*, 596-614 e514. 10.1016/j.cell.2021.01.002.
73. Bradner, J.E., Hnisz, D., and Young, R.A. (2017). Transcriptional Addiction in Cancer. Cell *168*, 629-643. 10.1016/j.cell.2016.12.013.
74. Massalha, H., Bahar Halpern, K., Abu-Gazala, S., Jana, T., Massasa, E.E., Moor, A.E., Buchauer, L., Rozenberg, M., Pikarsky, E., Amit, I., et al. (2020). A single cell atlas of the human liver tumor microenvironment. Mol Syst Biol *16*, e9682. 10.15252/msb.20209682.
75. Rambow, F., Rogiers, A., Marin-Bejar, O., Aibar, S., Femel, J., Dewaele, M., Karras, P., Brown, D., Chang, Y.H., Debiec-Rychter, M., et al. (2018). Toward Minimal Residual Disease-Directed Therapy in Melanoma. Cell *174*, 843-855 e819. 10.1016/j.cell.2018.06.025.
76. Oren, Y., Tsabar, M., Cuoco, M.S., Amir-Zilberstein, L., Cabanos, H.F., Hutter, J.C., Hu, B., Thakore, P.I., Tabaka, M., Fulco, C.P., et al. (2021). Cycling cancer persister cells arise from lineages with distinct programs. Nature *596*, 576-582. 10.1038/s41586-021-03796-6.